\documentclass[12pt]{iopart}
\usepackage[latin1]{inputenc}
\usepackage[english]{babel}
\usepackage{iopams}
\usepackage{setstack}
\usepackage{graphicx}

\begin{document}
\title[Time at which a Brownian motion is maximum before its first passage]{Distribution of the time at which the deviation of a Brownian motion is maximum before its 
first-passage time} 
\author{Julien Randon-Furling and Satya N Majumdar}
\address{Laboratoire de Physique Th\'eorique et Mod\`eles Statistiques, Universit\'e Paris-Sud, CNRS UMR 8626,  91405 Orsay Cedex, France}
\ead{julien.randon-furling@lptms.u-psud.fr}

\begin{abstract} 
We calculate analytically the probability density $P(t_m)$ of the time $t_m$ at which
a continuous-time Brownian motion (with and without drift) attains its maximum before passing 
through the origin for the first time. We also compute the joint probability density $P(M,t_m)$
of the maximum $M$ and $t_m$. In the 
driftless case, we find that $P(t_m)$ has power-law tails: $P(t_m)\sim t_m^{-3/2}$ for 
large $t_m$ and $P(t_m)\sim t_m^{-1/2}$ for small $t_m$. In presence of a drift towards the 
origin, $P(t_m)$
decays exponentially for large $t_m$. The results from numerical simulations are
in excellent agreement with our analytical predictions.

\end{abstract}
\maketitle

\section*{Introduction}

In this paper, we derive the probability distribution of a random variable
associated with a Brownian motion, namely the time
at which a Brownian motion attains its maximum value before it crosses
the origin for the first time. This random variable appears    
quite naturally in different problems such as in 
queueing theory and in the evolution of stock prices in finance.
 
Let us first consider, for example,
a single-server discrete-time queueing process, modelled as a simple
random walk~\cite{K,Asmussen} 
via:\[l_n=l_{n-1}+\xi _n,\] where $l_n$ is the length of the queue at time $n$ and $\xi 
_n$'s are independent and identically distributed random variables each
taking values $+1$ with probability $p$ (signifying the arrival of a new customer), $-1$
with probability $q$ (indicating the departure of an already served customer) or $0$
with probability $(1-p-q)$. In the queueing language, this is referred to as
the Geo/Geo/1 queue~\cite{Asmussen}.   

\begin{figure}[!ht]
\begin{center}
\includegraphics{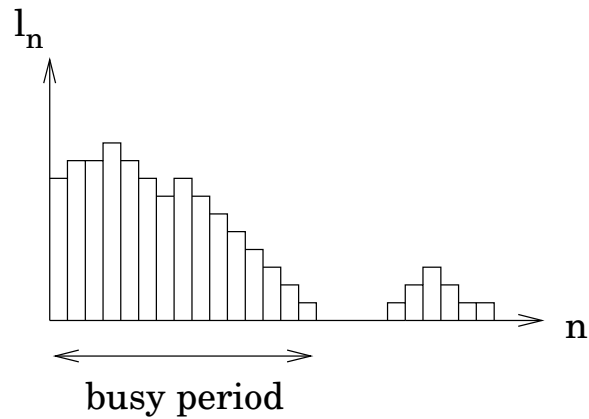}\caption{Queue with busy period.}
\end{center}
\end{figure}

Given $l_0$, one calls \textit{busy period} the period at the end of which the queue  
becomes empty for the first time (see Fig. 1): during such a period, the server 
always has some customers to serve. 
It is then natural to enquire about the time at which the queue is at its longest during 
the busy period. In the random walk model where the queue length
$l_n$ is the position of the walker at time step $n$, this amounts to investigating 
the time 
at which the position of the walker (initially positive) is farthest from
the origin before it crosses the origin for the first time. 

Another area where the same variable appears quite naturally is in the
evolution of stock prices in finance. The evolution of
a stock price $S_n$ with time $n$ is often modelled by the
exponential of a random walk~\cite{Wil,Yor}. Starting from its initial value $S_0$ 
the price evolves with time stochastically. A natural question for an agent holding 
this stock is: what is
the suitable time for selling this stock? If the stock price goes below a
threshold, say $R$, it is too risky to wait any longer. Thus an agent can wait at most
up to the time at which the ratio $S_n/R$ crosses the level $1$ from above.
Within this time, the ratio $S_n/R$  will achieve its maximum at some
intermediate time which is clearly the best time to sell the stock. 
Assuming that the random variable $l_n =\ln (S_n/R)$ performs a random walk
starting from its initial value $l_0=\ln (S_0/R)>0$, one then wants to
calculate the probability distribution of the time at which the random
walk is farthest from the origin till its first-passage time through the origin,
i.e., till the time at which $S_n/R=1$, i.e., $\ln (S_n/R)=0$ indicating the 
first-passage through the origin. 
     
In this paper, we will consider a further simplified case, namely a continuous-time
Brownian motion as opposed to the discrete-time random walk in the above two problems.
For a continuous-time Brownian motion we calculate explicitly, using path-integral
methods, the 
probability
density of the time $t_m$ at which a Brownian motion $x(t)$ (starting from $x(0)=x_0>0$)
is farthest from the origin before it crosses the origin for the first time.
Algorithmically speaking,
for each sample of the Brownian motion starting at $x_0$ we
stop when it crosses the origin for the first time say at time $t=t_f$ and locate the 
time $0\le t_m < t_f$ 
at which the Brownian motion achieves its maximum value. Note that both $t_f$ (the 
first-passage time) and $t_m$ varies from one sample to another. We repeat it many times 
and then construct a histogram of the $t_m$'s which gives its probability density function 
$P(t_m)$. Even though the discrete-time problem is more relevant, we expect the 
continuous-time
result to provide the right asymptotics for the discrete problem. As we will see 
below, the continuous-time problem, though still non-trivial, is easier to handle 
analytically.  

Note that for a Brownian motion or a Brownian bridge over a {\it fixed} time interval 
$[0,T]$, the probability density $P(\tau, T)$ of the time $\tau$ at which the process 
attains its maximum is well known~\cite{Fel}.  
For example, for a zero-drift Brownian motion over $[0,T]$ starting at the origin, the 
probability density 
$P(\tau,T)=\frac{1}{T}g(\tau/T)$ where $g(x)= 1/[\pi \sqrt{x(1-x)}]$ for $x\in 
[0,1]$~\cite{Fel}. On the other hand, for a Brownian bridge over the
fixed interval $[0,T]$ and starting at the 
origin,
the probability density of $\tau$ is uniform, $P(\tau,T)=1/T$ for $0\le \tau\le 
T$~\cite{Fel}. In contrast, in our case, the Brownian motion is not over 
a fixed time interval, but rather over a variable time interval $[0,t_f]$ where
the upper edge $t_f$ is the first-passage time which itself
is a random variable~\cite{Red} and hence varies from sample to sample.
 
The statistical properties of the functionals (such as the area, the maximum etc.) of a 
Brownian motion or its variants 
(such as a bridge, excursion, meander etc.) over a fixed time interval have many
applications in physics, graph theory, computer science and they have been 
studied extensively~(for recent reviews on Brownian functionals see~\cite{CDT,SM}).
In particular, the area under a Brownian excursion or meander has found
many recent applications in problems as diverse as fluctuating interfaces~\cite{SMAC}, graph enumeration~\cite{SvJ}, lengths of internal paths in rooted planar trees~\cite{SM,Tak} or cost functions in data storage via the ``linear probing with hashing'' algorithm~\cite{SM,FPV}. 
Similarly,   
the statistical properties of functionals of Brownian motion restricted up to its 
first-passage time (usually referred to as  `first-passage functionals') also
have various applications, and have appeared recently in many  
different contexts~\cite{K,KM,KMM}, including the computation of the time period of oscillation of an undamped particle in a random potential~\cite{DM} and the determination of the distribution of the lifetime of a comet in the solar system~\cite{SM,Ham}. The probability density of the area
swept by an initially positive Brownian motion till its first-passage time was computed exactly in \cite{KM}, with an application to queuing theory.
In this paper, our focus is on the random variable
$t_m$ which, though not quite a functional in the strict sense, is an important
random variable associated with such a Brownian motion
restricted up to its first-passage time.
\begin{figure}[!h]
 \begin{center}
 \includegraphics[width=10cm,height=7cm]{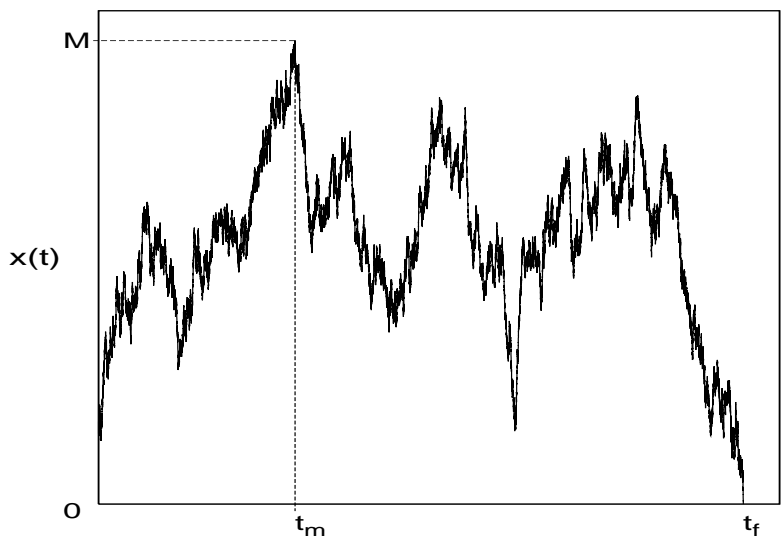}
\caption{Maximum before the first-passage through the origin for the zero-drift case.} 
\end{center}
\end{figure}

In \cite{KM}, the authors also computed directly the probability density $P(M)$ of the 
maximum $M$ of a 
Brownian motion (starting at $x_0>0$) before its first-passage time through the origin, 
via a ``backward'' 
Fokker-Planck method and showed that it has a power law behavior $P(M)=x_0/M^2$ 
where $M\ge x_0$. In this paper, we extend this work using a path decomposition method 
that allows us to obtain the joint probability density $P(M,t_m)$ of the maximum $M$
and the time $t_m$ at which the maximum occurs before the first-passage time. By 
integrating
over $M$, we then get the `marginal' $P(t_m)$, i.e. the probability density of $t_m$.
We calculate $P(t_m)$ explicitly both for a driftless and drifted Brownian motion.
We also compare the results of numerical simulations to our analytical predictions
and find excellent agreement.

\section{Driftless Case} 

We consider a continuous-time Brownian motion evolving via $dx/dt=\xi(t)$, where
$\xi(t)$ is a white noise with $\langle \xi(t)\rangle =0$ and $\langle \xi(t)\xi(t')\rangle 
=\delta(t-t')$.   
We start by recalling the quick derivation of $P(M)$ given in 
\cite{KM}:

\vspace{\baselineskip}

Let $q(x)$ be the probability that a Brownian particle starting from $x \in 
[0,M]$ exits the interval for the first time through $0$, \textit{i.e.}, the probability that 
the maximum before the first-passage time is less than or equal to $M$. Writing $\phi_{\Delta t}(\Delta x)$ for the 
distribution 
function of a Brownian displacement in the time interval $\Delta t$, we have: \begin{equation}q(x) = 
\int q(x+\Delta x) \phi_{\Delta t}(\Delta x) \,d\Delta x.\end{equation}
 Expanding $q(x+\Delta x)$ for 
small values of $\Delta x$, and using the fact that in the absence of drift the mean value 
of $\Delta x$ is $0$, one finds that $q$ satisfies: 
\begin{equation}\frac{d^2 
q}{dx^2}=0,\:q(0)=1,\:q(M)=0,\label{eqq}\end{equation} 
whose solution is: \begin{equation} 
q(x)=1-\frac{x}{M}.\label{q}\end{equation} As mentioned in its definition, it can easily be seen that $q(x)$ also corresponds to the probability that the maximum before the first-passage time is less than or equal to $M$; therefore, differentiating Eq.~\ref{q} with respect to $M$ gives the probability density of $M$: 
\begin{equation}P(M)=\frac{x}{M^2}.\label{PM}
\end{equation}

To compute the joint probability density $P(M,t_m)$ we proceed
as follows. We first assume that the maximum occurs at $t_m$ and then we split 
the Brownian path into two parts (before/after $t_m$, as shown in 
Fig.~3) and determine the weights of a path's left-hand side and right-hand side 
separately. Note that due to the \textit{Markovian} property of the Brownain path, 
once the position  
of the walker is specified at $t_m$, the weights of the left and the right parts
become completely \textit{independent} and the total weight is just proportional
to the product of the weights of the two separate parts. For the left part, we have
a process that propagates from $x_0$ at $t=0$ to $M$ at $t=t_m$ without crossing
the level $M$ in $[0,t_m]$ (since $M$ is the maximum) and the level $0$ (the origin). For the 
right part,
the process propagates from the value $M$ at $t=t_m$ to $0$ at $t=t_f$ where
$t_f\ge t_m$ without crossing the level $M$ and the level $0$ 
in between. We need to be careful, however, because, as pointed out in \cite{SMAC}, a 
Brownian walker that crosses a given level once crosses it infinitely many times 
immediately after the first crossing. It is therefore impossible to enforce
the constraint $x(t_m)=M$ and simultaneously forcing the motion to
stay below $M$ before or after $t_m$ (for a lattice walk, one does not
have this problem since the lattice constant provides a natural cut-off). 
Following the method used in~\cite{SMAC}, we introduce a cut-off $\epsilon$ by 
imposing $x(t_m)=M-\varepsilon$
and 
consider all paths having a 
maximum less than or equal to $M$ and passing through $M-\varepsilon$ at $t=t_m$. We 
compute their weight and then let $\varepsilon$ go to $0$ eventually.

\begin{figure}[!h]
 \begin{center}
\includegraphics[width=10cm,height=7cm]{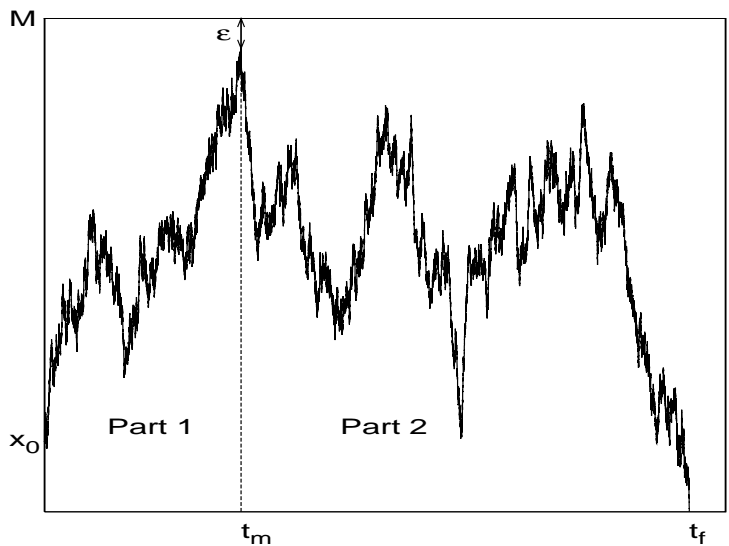} \caption{Splitting 
probabilities}
\end{center}\label{SpP}\end{figure} 

\noindent On the right side of $t=t_m$:  we 
have to determine the weight of a path that starts at $M-\varepsilon$ and exits for
the first time the interval 
$[0,M]$ through $0$. This is given by Eq.~\ref{q}:
\begin{equation}
 q(M-\varepsilon)=\frac{\varepsilon}{M}\label{rhs}
\end{equation}

\noindent On the left side of $t=t_m$: we use a path integral treatment with the Feynman-Kac 
formula (as in \cite{SMAC}) giving the weight of a path in terms of the propagator 
$\left\langle x_0 \left|\,e^{-\hat{H}t_m}\,\right| M-\varepsilon \right\rangle$, where 
$\hat{H}=-\frac{1}{2}\frac{\partial^2}{\partial x^2}+V(x)$ with $V(x)$ a square well having 
infinite barriers at $x=0$ and $x=M$ and $V(x)=0$ for $0<x<M$ (the infinite barriers at $x=0$, $x=M$ enforce the condition that the path can penetrate neither at $x=0$ nor at $x=M$). The normalized eigenfunctions
of $\hat{H}$ labelled by the integer $n=1,2,3,\ldots $ are $\psi_n(x)= 
\sqrt{\frac{2}{M}}\,\sin\left(\frac{n\pi x}{M}\right)$ with
the associated eigenvalues $E_n=n^2\pi^2/{2M^2}$. The eigenfunction
$\psi_n(x)$ vanishes at both ends $x=0$ and $x=M$ of the box. The propagator
can be easily evaluated in this eigenbasis,
$\left\langle x_0 \left|\,e^{-\hat{H}t_m}\,\right| M-\varepsilon \right\rangle
=\sum_{n=1}^{\infty} \psi_n(x_0)\psi_n(M-\varepsilon)\,e^{-E_n t_m}$ and one gets:
\begin{equation}
\fl \left\langle x_0 \left|\,e^{-\hat{H}t_m}\,\right| M-\varepsilon \right\rangle 
= \frac{2}{M}\sum_{n=1}^{\infty} \sin\left(\frac{n\pi x_0}{M}\right)
\sin\left(\frac{n\pi (M-\varepsilon)}{M}\right)\,e^{-\frac{n^2\pi^2}{2M^2}t_m}.
\label{propa1}
\end{equation}
In the limit when $\varepsilon \to 0$, we get to leading order
\begin{equation}
\fl \left\langle x_0 \left|\,e^{-\hat{H}t_m}\,\right| M-\varepsilon 
\right\rangle=\frac{2\pi}{M^2}\varepsilon 
\displaystyle\sum_{n=1}^{\infty}(-1)^{n+1}n\sin \left(\frac{n\pi x_0}{M}\right) 
e^{-\frac{n^2\pi^2}{2M^2}t_m} + \Or(\varepsilon^2).
\label{lhs}
\end{equation} 

\noindent Taking the product of Eqs.~\ref{rhs} and \ref{lhs}, we get the total
weight of the path, to leading order in small $\varepsilon$,
\begin{equation}
P(M,t_m; \varepsilon) \propto \varepsilon^2 \frac{\pi}{M^3}\,
\sum_{n=1}^{\infty}(-1)^{n+1}\,n\,\sin 
\left(\frac{n\pi x_0}{M}\right) e^{-\frac{n^2\pi^2}{2M^2}t_m}.
\label{pjoint}
\end{equation}
The proportionality constant is set by using the normalization constant:
\[\int_{x_0}^{\infty} dM \int_0^{\infty} dt_m P(M,t_m;\varepsilon)=1.\]
It is easy to show that the proportionality constant   
$A(\varepsilon)=\varepsilon^{-2}$. Thus, in the limit $\varepsilon\to 0$, 
we finally obtain 
\begin{equation}
P\left(M,t_m\right)=\frac{\pi}{M^3}
\displaystyle\sum_{n=1}^{\infty}(-1)^{n+1}\,n\,\sin 
\left(\frac{n\pi x_0}{M}\right) e^{-\frac{n^2\pi^2}{2M^2}t_m}\label{JP} 
\end{equation}
As a first check, let us show that  
$\int_{0}^{\infty}\,dt_m\:P\left(M,t_m\right)= x_0/M^2$, thus recovering
the marginal $P(M)$ of the maximum in 
Eq.~\ref{PM}. Integrating over $t_m$, we get
\begin{equation}
P(M)=\frac{2}{\pi M}\, \sum_{n=1}^{\infty} \frac{(-1)^{n-1}}{n}\, \sin\left(\frac{n\pi 
x_0}{M}\right)=\frac{x_0}{M^2}
\label{PMM}
\end{equation}
where the last identity can be found (and derived easily) in ~\cite{GR}.

Finally, from Eq.~\ref{JP} an integration over $M$ (note that $M$ varies from $x_0$
to $\infty$) yields the desired marginal $P(t_m)$:
\begin{eqnarray}
 P(t_m) & = & \pi 
\displaystyle\int_{x_0}^{\infty}\,\frac{dM}{M^3}\:
\displaystyle\sum_{n=1}^{\infty}(-1)^{n+1}\,n\,\sin 
\left(\frac{n\pi x_0}{M}\right) e^{-\frac{n^2\pi^2}{2M^2}t_m} \nonumber \\
&=&\frac{1}{\pi t_m} \displaystyle\sum_{n=1}^{\infty}
\frac{(-1)^{n+1}}{n}\displaystyle\int_{0}^{n\pi}\,du\:
\cos (u) e^{-\frac{u^2}{2x_0^2}t_m}. \label{PTs}
\end{eqnarray}
The sum in Eq.~\ref{PTs} can be expressed in terms of a known special function
and we get
\begin{equation}
P\left(t_m\right)= \frac{1}{2\pi t_m}\left[\pi - \int_{0}^{\pi}
\vartheta_4 \left(\frac{y}{2},e^{-y^2\frac{t_m}{2 x_0^2} }\right)dy\right]\label{PT}
\end{equation} 
where $\vartheta_4(z,q)$ is the fourth of Jacobi's Theta functions (\cite{AbSe}).
Subsequently one can obtain the large and small $t_m$ asymptotics of $P(t_m)$ from the
exact expression in Eq.~\ref{PT}. 

\subsubsection*{Large-$t_m$ asymptote:} We first consider the case when
$t_m\gg x_0^2$.
Changing variables in Eq.~\ref{PTs} through 
$z=\sqrt{\frac{t_m}{2x_0^2}}\,u$ and letting $z \rightarrow 0$ gives for $t_m\gg x_0^2$: 
\begin{equation} 
P(t_m) \approx \frac{x_0 \log 
2}{t_m^{3/2}}\sqrt{\frac{1}{2\pi}}
\label{PTI} \end{equation}

\subsubsection*{Small-$t_m$ asymptote:}
In the opposite limit $t_m\ll x_0^2$,
we start from Eq.~\ref{JP}  and first take a Laplace transform:
\begin{eqnarray*}
\displaystyle\int_{0}^{\infty}\,dt_m\:
e^{-st_m}P\left(M,t_m\right)&=\frac{2}{\pi M}
\displaystyle\sum_{n=1}^{\infty}(-1)^{n+1}
\frac{n}{(n^2+\frac{2M^2s}{\pi^2})}\,\sin \left(\frac{n\pi x_0}{M}\right)\\
&= \frac{\sinh (x_0\sqrt{2s})}{M\sinh (M\sqrt{2s})},\\
\end{eqnarray*}
where the sum of the series can be found in \cite{GR}.
Letting $s$ become much larger than $x_0^{-2}$ and $M^{-2}$, we obtain:
\[\displaystyle\int_{0}^{\infty}\,dt_m\:e^{-st_m}P\left(M,t_m\right) 
\approx \frac{e^{-\sqrt{2s}(M-x_0)}}{M},\]
which, after the Laplace inversion (\cite{Ba}) yields:
\begin{equation}P\left(M,t_m\right)\approx 
\frac{t_m^{-3/2}}{\sqrt{2\pi}}\frac{\left(M-x_0\right)}{M}
e^{-\frac{\left(M-x_0\right)^2}{2t_m}}.\label{JPSA}
\end{equation}
Integrating over $M$ gives for $t_m\ll x_0^2$:
\begin{equation}
P(t_m)\approx \frac{1}{x_0\sqrt{2\pi t_m}}\label{PT0}
\end{equation}

Thus, $P(t_m)$ has power law behavior at both large and small tails. For 
large $t_m$, the probability density falls off as $P(t_m)\sim t_m^{-3/2}$, whereas for small 
$t_m$ it diverges as $P(t_m)\sim t_m^{-1/2}$. 
The exact analytical form of $P(t_m)$ and its asymptotes from Eqs.~\ref{PT}, 
\ref{PTI}, \ref{PT0} are 
plotted (using Mathematica) in Fig.~4 together with the 
points obtained from the numerical simulation (with 1,000,000 realisations). They are
in good agreement with each other. 
\begin{figure}[h]
\begin{center}
\includegraphics[scale=0.65]{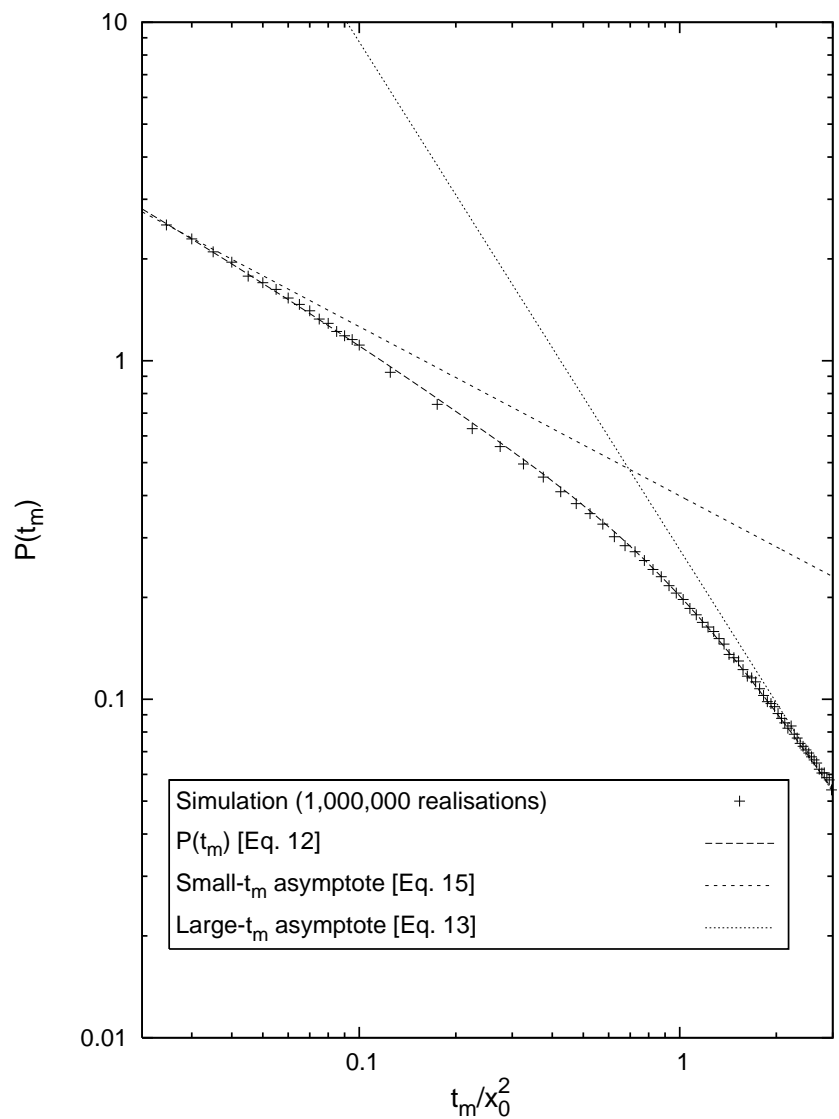}\caption{The probability density $P(t_m)$ and
its asymptotes for the driftless case. In the simulation, $x_0$ was set to $1$.}
\end{center}\label{DrLP}
\end{figure}
\cleardoublepage

\section{In the Presence of a Negative Drift} 

We now consider a Brownian motion in the presence of a drift $\mu$. For $\mu>0$, it is 
clear
that the walker will excape to $\infty$ with a nonzero probability. This means that
with a finite probability $t_m\to \infty$. Therefore this case is not much of interest
in the present context. Instead, we focus here on the opposite case where the drift 
is towards the origin, i.e., $\mu<0$.
The Langevin equation describing the motion 
becomes: \[\frac{dx}{dt}=-|\mu| +\xi (t),\] where $\xi (t)$ is the Gaussian white noise
with $\langle \xi(t)\rangle=0$ and $\langle \xi(t)\xi(t')\rangle=\delta(t-t')$. We use 
the same strategy as in the driftless case, \textit{i.e.}, splitting the motion into two 
independant 
parts (before and after $t_m$) and introducing a small cut-off $\varepsilon$. 

\vspace{\baselineskip}

\noindent On the right-hand side: Letting, as in the driftless case, $q(x)$ be the probability that a Brownian particle starting from $x \in 
[0,M]$ exits the interval for the first time through~$0$, we have as before:
\begin{equation}q(x) = 
\int q(x+\Delta x) \phi_{\Delta t}(\Delta x) \,d\Delta x.\end{equation} In the presence of a drift, the mean value of $\Delta x$ is no longer $0$ and one can easily show that the analogue of Eq.~\ref{eqq} now reads: \begin{equation} q''(x)-2|\mu| q(x) = 0,\qquad q(0)=1,\:q(M)=0.\end{equation}
The solution is: \begin{equation} q(x) = \frac{\sinh (|\mu|(M- x))}{\sinh (|\mu| M)}e^{|\mu| x}\label{dq},\end{equation} and so we have:
\begin{equation}
 q(M-\varepsilon)= \frac{\sinh (|\mu|\varepsilon)}{\sinh (|\mu| M)}e^{|\mu| (M-\varepsilon)}.\label{qd}
\end{equation}
As in the driftless case, the probability density of $M$ can be obtained by differentiation of Eq.~\ref{dq} with respect to $M$, as was done in \cite{KM}:
\begin{equation} P_d(M)=\frac{|\mu|\sinh (| \mu| x)}{\sinh^2 (| \mu | M)}e^{| \mu| x},\end{equation} 
where we have added the subscript ``d'' to indicate that the density corresponds to 
the drifted case. 

\vspace{\baselineskip}

\noindent On the left-hand side: We use the same path integral method as in the driftless case. The weight of a path is now proportional to: 
\begin{eqnarray} 
\fl \exp \left[ -\frac{1}{2}\int_0^{t_m}d\tau\:\left(\frac{dx}{d\tau}+|\mu|\right)^2\right] \nonumber \\=  \exp \left[ -\frac{|\mu|^2}{2}t_m-|\mu|\int_0^{t_m}d\tau\:\frac{dx}{d\tau}\right] \exp \left[ -\frac{1}{2}\int_0^{t_m}d\tau\:\left(\frac{dx}{d\tau}\right)^2\right].\label{fact}
\end{eqnarray}
The position of the Brownian particle at $t=0$ and $t=t_m$ is known, so we can substitute $(M-\varepsilon)-x_0$ for $\int_0^{t_m}d\tau\:\frac{dx}{d\tau}$ in the first exponential factor on the right-hand side of Eq.~\ref{fact}.

The propagator for the drifted case will therefore be equal to that for the driftless case (given in Eq.~\ref{propa1}) multiplied by the factor $\exp \left[ |\mu| x_0-\frac{| \mu|^2}{2}t_m-|\mu|(M-\varepsilon)\right] $, and will be given by:
\begin{equation}
\fl \exp \left[ |\mu| x_0-\frac{| \mu|^2}{2}t_m-|\mu|(M-\varepsilon)\right] \frac{2}{M}\sum_{n=1}^{\infty} \sin\left(\frac{n\pi x_0}{M}\right) \sin\left(\frac{n\pi (M-\varepsilon)}{M}\right)\,e^{-\frac{n^2\pi^2}{2M^2}t_m}.\label{ld}
\end{equation}

\vspace{1em}

As in the driftless case, we multiply the weights of the left and right side of $t_m$ 
derived above (Eqs.~\ref{qd} and \ref{ld}), and take the $\varepsilon 
\rightarrow 0$ limit to obtain: 
\begin{equation} P_d(M,t_m)=\frac{| \mu| M e^{|\mu| 
x_0-\frac{| \mu|^2}{2}t_m}}{\sinh \left( |\mu| M\right)} P(M,t_m)
\label{JPd} 
\end{equation} 
where $P(M,t_m)$ is the joint density for the driftless
case given in Eq. (\ref{JP}). Once again, by integrating over $t_m$, one can
recover the marginal probability density of the maximum $P_d(M)$ derived
originally in \cite{KM}. On the other hand, integrating over $M$
gives the marginal $P_d(t_m)$. We were not able to derive a compact
expression for $P_d(t_m)$ as in the driftless case, though the
asymptotes of $P_d(t_m)$ can be derived explicitly as shown below.  

\subsubsection*{Small-$t_m$ asymptote:} 
From Eq.~\ref{JPd}, we can derive very quickly the behaviour of $P_d(t_m)$ when $t_m \ll 
x_0^2$. Substituting in Eq.~\ref{JPd}
the asymptotic result for the driftless case from 
Eq.~\ref{JPSA} and integrating over $M$ we get: 
\begin{equation}P_d(t_m)\sim\frac{|\mu|e^{|\mu|x_0 - \frac{| \mu|^2}{2}t_m}}{\sinh 
(|\mu |x_0) \sqrt{2\pi t_m}} \label{PdSA} \end{equation}
Thus for small $t_m$, $P_d(t_m)$ diverges as $t_m^{-1/2}$, as in the driftless case.

\subsubsection*{Large-$t_m$ asymptote:} To study the behaviour of $P_d(t_m)$ when $t_m \gg 
x_0^2$, we start from the following expression for $P_d(t_m)$: \begin{eqnarray}
 P_d(t_m) =  \displaystyle\int_{x_0}^{\infty}\,dM\:P_d(M,t_m)\nonumber\\
= \displaystyle\int_{x_0}^{\infty}\,dM\:\frac{| \mu| M e^{|\mu| x_0-\frac{| \mu|^2}{2}t_m}}{\sinh \left( \mu M\right)} P(M,t_m)\nonumber\\
= \displaystyle\int_{x_0}^{\infty}\,
dM\:\frac{|\mu|\pi e^{|\mu| x_0-\frac{| \mu|^2}{2}t_m}}
{\sinh (|\mu| M)M^2}\displaystyle\sum_{n=1}^{\infty}(-1)^{n+1}n\sin 
\left(\frac{n\pi x_0}{M}\right) e^{-\frac{n^2\pi^2}{2M^2}t_m}.
\label{pdtm}
\end{eqnarray}
The series in Eq.~\ref{pdtm} is dominated by the first term ($n=1$) for large~$t_m$. Hence,
retaining only the $n=1$ term
and making a change of variable $y=1/M$ in the integral, we get:
\[\pi |\mu| e^{|\mu|x_0 - \frac{| \mu|^2}{2}t_m}
\displaystyle\int_{0}^{\frac{1}{x_0}}\, 
dy\: \frac{\sin(\pi x_0 y)}{\sinh (|\mu| / y)}
e^{-\frac{\pi^2y^2 t_m}{2}}.\]
For large $t_m$, the most important contribution to the integral comes from the small
$y$ regime. 
Expanding the $\sin$ and $\sinh$ functions and keeping only the leading order term
reduces the integral to: 
\[2\pi^2 x_0 |\mu| e^{|\mu|x_0 - \frac{| 
\mu|^2}{2}t_m}\displaystyle\int_{0}^{\frac{1}{x_0}}\,dy\:y\,e^{-t_m\left(\frac{\pi^2}{2} 
y^2+\frac{|\mu|}{y t_m}\right) }.\] 
Letting $h(y)=\frac{\pi^2}{2}y^2+\frac{|\mu|}{y t_m}$, we 
next use the saddle point method to obtain the leading term via minimizing the function $h$
and get:
\begin{equation}P_d(t_m)\sim \left[ 
2\sqrt{\frac{2}{3}}\pi^\frac{5}{6}x_0|\mu|^\frac{4}{3}e^{|\mu|x_0}\right]\, 
t_m^{-5/6}\, e^{-\frac{|\mu|^2}{2}t_m-\frac{3}{2}\left( |\mu|\pi\right) 
^{2/3}t_m^{1/3}}\label{PdLA}\end{equation} \vspace{2em}

Thus, as expected, the density $P_d(t_m)$ has an exponential decay for large $t_m$
in presence of a negative drift.
\noindent Figure~5 shows a plot of the asymptotes (Eq.~\ref{PdSA} and \ref{PdLA}) 
together with the data from numerical simulation (1,000,000 realisations with 
$|\mu|=0.1$). 
\begin{figure}[h] \begin{center}
\includegraphics[width=11cm,height=12.5cm]{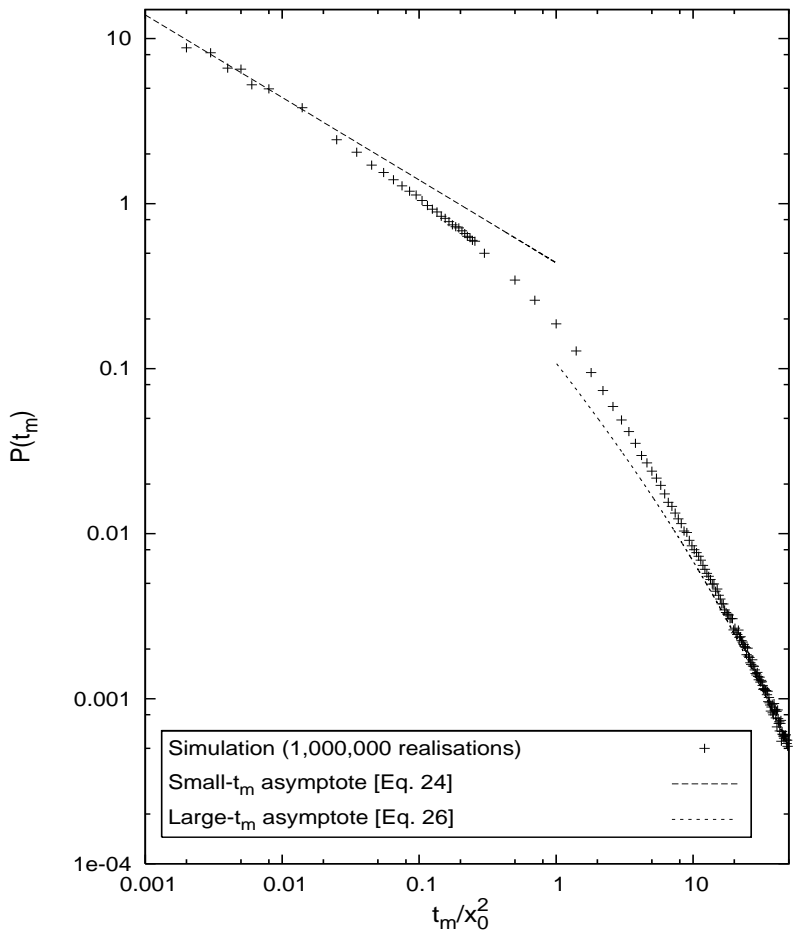} \caption{The simulated probability density $P(t_m)$  and its asymptotes in the presence of a drift towards the origin ($|\mu|=0.1$). In the simulation, $x_0$ was set to $1$.}
\end{center}\label{DrP}\end{figure} \cleardoublepage

\section{Summary and Conclusion} In summary, we have obtained an exact expression for the 
probability density 
of the time at which a Brownian motion attains its maximum before passing through the 
origin for the first time, and studied the tails of this probability density both for 
the driftless 
and for the drifted Brownian motion. This was done by first computing the joint 
distribution 
$P(M,t_m)$ of the maximum $M$ attained and the time $t_m$ at which it is attained.
In the context of the queuing theory, the result that $P(t_m)$ decreases monotically with increasing $t_m$ suggests that the beginning of a busy period is more likely to be the time at which a queue is at its longest.

It would be interesting to derive the explicit results, obtained here by the path integral method, from the general theory of filtrations in Brownian motion developed recently in \cite{MY,NY}.

It would also be of interest to extend this calculation to the discrete-time random walk
case which remains a real challenge.
\ack We thank A. Comtet, M.J. Kearney and P.L. Krapivsky 
for useful discussions. We also thank M. Yor for useful discussions and for pointing out references \cite{MY,NY}.


\section*{References}
\begin{thebibliography}{10}
\bibitem{K} Kearney M J 2004,{\it J. Phys. A. Math. Gen.} {\bf 37}, 8421.

\bibitem{Asmussen} Asmussen S 2003, {\it Applied Probability and Queues} 2nd edn (New York: Springer). 

\bibitem{Wil} Williams R J 2006, {\it {I}ntroduction to the {M}athematics of {F}inance} (AMS).

\bibitem{Yor} Yor M 2000, {\it Exponential Functionals of Brownian Motion and Related
Topics} (Berlin: Springer); see also Comtet A, Monthus C and Yor M 1998, J. Appl. Prob.
{\bf 35}, 255. 

\bibitem{Fel} Feller W 1968, {\it {A}n {I}ntroduction to {P}robability {T}heory and its 
{A}pplications} (New York: Wiley).

\bibitem{Red} Redner S 2001, {\it A Guide to First-Passage Processes} (Cambridge: Cambridge 
University Press).

\bibitem{CDT} Comtet A, Desbois J and Texier C 2005,{\it J. Phys. A. Math. Gen.} {\bf 38}, R341.

\bibitem{SM} For a short review on Brownian functionals and their applications see  
Majumdar S N 2005, {\it Current Science}, {\bf 89}, 2075; also available at 
http://xxx.arXiv.org/cond-mat/0510064. 

\bibitem{SMAC} Majumdar S N and Comtet A 2005,{\it J. Stat. Phys.} {\bf 119}, 777; 
2004 Phys. Rev. Lett. {\bf 92}, 225501.

\bibitem{SvJ} For an extensive review on the area under Brownian motion 
and its variants, see Janson S 2007, {\it Prob. Surveys} {\bf 4}, 80.

\bibitem{Tak} Tak\'{a}cs L 1991, {\it Adv. Appl. Prob.} {\bf 23}, 557; 1995, {\it J. Appl. Prob.} {\bf 32}, 375.

\bibitem{FPV} Flajolet P, Poblete P and Viola A 1998, {\it Algorithmica} {\bf 22}, 490.

\bibitem{Ham} Hammersley J M 1961, {\it Proc. 4th Berkeley Symp. on Math. Stat. and Prob.} vol.~3 (Berkeley: Univ. Cal. Press), pp.~17.

\bibitem{DM} Dean D S and Majumdar S N 2001, {\it J. Phys. A. Math. Gen.} {\bf 34}, L697.

\bibitem{KM} Kearney M J and Majumdar S N 2005, {\it J. Phys. A: Math. Gen.} {\bf 38}, 4097.

\bibitem{KMM} Kearney M J, Majumdar S N and Martin R.J. 2007, arXiv:0706.2038.


\bibitem{GR} Gradshteyn I S and Ryzhik I M 2000, {\it Table of Integrals, Series and Products}
6th edn., (New York: Academic Press).

\bibitem{Ba} Bateman H 1954, {\it Tables of Integral Transforms} (McGraw-Hill).

\bibitem{AbSe} Abramowitz M and Stegun I A 1973, {\it Handbook of Mathematical Functions} 
(Dover). 

\bibitem{MY} Mansuy R and Yor M 2006, {\it Lecture Notes in Mathematics} 1873 (Berlin: Springer Verlag).

\bibitem{NY} Nikeghbali A and Yor M 2006, {\it Ill. J. Math.} {\bf 50}, 791.

\end{thebibliography}
\end{document}